\documentclass[runningheads]{llncs}
\usepackage[T1]{fontenc}
\usepackage{amsmath,amssymb}
\usepackage{booktabs}
\usepackage{siunitx}
\usepackage{multirow}
\usepackage{hyperref}
\usepackage{orcidlink}
\usepackage{tikz}
\hypersetup{
colorlinks=true,
linkcolor=black,
citecolor=black
}
\usepackage{graphicx}
\usepackage[utf8]{inputenc}
\usepackage{newunicodechar}

\newunicodechar{（}{(}\newunicodechar{）}{)}
\begin{document}
\title{MisEdu-RAG: A Misconception-Aware Dual-Hypergraph RAG for Novice Math Teachers}

\titlerunning{A Misconception-Aware Dual-Hypergraph RAG for Novice Math Teachers}
%
%\titlerunning{Abbreviated paper title}
% If the paper title is too long for the running head, you can set
% an abbreviated paper title here
%
%\author{First Author\inst{1}\orcidID{0000-1111-2222-3333} \and
%Second Author\inst{2,3}\orcidID{1111-2222-3333-4444} \and
%Third Author\inst{3}\orcidID{2222--3333-4444-5555}}
%
%\authorrunning{F. Author et al.}
% First names are abbreviated in the running head.
% If there are more than two authors, 'et al.' is used.
%
%\institute{Princeton University, Princeton NJ 08544, USA \and
%Springer Heidelberg, Tiergartenstr. 17, 69121 Heidelberg, Germany
%\email{lncs@springer.com}\\
%\url{http://www.springer.com/gp/computer-science/lncs} \and
%ABC Institute, Rupert-Karls-University Heidelberg, Heidelberg, Germany\\
%\email{\{abc,lncs\}@uni-heidelberg.de}}
%\maketitle

%\author{Anonymous Author(s)}
%\authorrunning{Anonymous et al.}
%\institute{Anonymous Institution(s)}
\author{
Zhihan Guo\inst{1}$^{~\orcidlink{0009-0004-7638-6779}}$ \and
Yuting Lu\inst{1}$^{~\orcidlink{0009-0006-2745-0222}}$ \and
Jionghao Lin\inst{1\thanks{Corresponding author.},2,3}$^{~\orcidlink{0000-0003-3320-3907}}$
}

\authorrunning{Z. Guo et al.}

\institute{
The University of Hong Kong, Hong Kong SAR, China\\
\email{zhihang330@connect.hku.hk}\\
\email{yutinglu@connect.hku.hk}\\
\email{jionghao@hku.hk}
\and
Carnegie Mellon University, Pittsburgh PA 15213, USA  \and     Monash University, Clayton VIC 3800, Australia
}

\maketitle
\begin{abstract}
Novice math teachers often encounter students' mistakes that are difficult to diagnose and remediate. Misconceptions are especially challenging because teachers must explain what went wrong and how to solve them. Although many existing large language model (LLM) platforms can assist in generating instructional feedback, these LLMs loosely connect pedagogical knowledge and student mistakes, which might make the guidance less actionable for teachers. To address this gap, we propose MisEdu-RAG, a dual-hypergraph-based retrieval-augmented generation (RAG) framework that organizes pedagogical knowledge as a concept hypergraph and real student mistake cases as an instance hypergraph. Given a query, MisEdu-RAG performs a two-stage retrieval to gather connected evidence from both layers and generates a response grounded in the retrieved cases and pedagogical principles. We evaluate on \textit{MisstepMath}, a dataset of math mistakes paired with teacher solutions, as a benchmark for misconception-aware retrieval and response generation across topics and error types. Evaluation results on \textit{MisstepMath} show that, compared with baseline models, MisEdu-RAG improves token-F1 by 10.95\% and yields up to 15.3\% higher five-dimension response quality, with the largest gains on \textit{Diversity} and \textit{Empowerment}. To verify its applicability in practical use, we further conduct a pilot study through a questionnaire survey of 221 teachers and interviews with 6 novices. The findings suggest that MisEdu-RAG provides diagnosis results and concrete teaching moves for high-demand misconception scenarios. Overall, MisEdu-RAG demonstrates strong potential for scalable teacher training and AI-assisted instruction for misconception handling. Our code is available on GitHub: \url{https://github.com/GEMLab-HKU/MisEdu-RAG}.

\end{abstract}
%

% max: 14 pages

\section{Introduction}

Recent advancements in Large Language Models (LLMs) have shown promise to support teacher professional development (PD) training \cite{wang2024large}, especially for novice teachers, who are typically in their first few years of teaching \cite{moosapoor2023new}. Within teacher PD, supporting novice teachers is especially important, as they often struggle to translate recognized student misconceptions into effective instructional responses \cite{moosapoor2023new}. However, providing misconception-focused support requires reliable and pedagogically grounded guidance, while LLM outputs can be inconsistent and may contain hallucinations \cite{ji2023survey}. Retrieval-Augmented Generation (RAG) improves reliability by retrieving an external reference corpus to support generation. It attaches the retrieved evidence to the LLM input so the model responds based on that evidence rather than relying only on its internal knowledge \cite{lewis2020retrieval,li2025retrieval}. However, misconception-aware support often requires linking pedagogical principles with similar student cases and integrating evidence distributed across multiple sources, which flat passage retrieval may not capture well. Graph-based retrieval explicitly models relations among concepts and cases, enabling retrieval of connected evidence for more coherent aggregation \cite{edge2024local,feng2025hyper}.

To address this gap, we propose \textbf{MisEdu-RAG}, a dual-hypergraph RAG framework for misconception-aware teacher training. Effective misconception support requires both concept-level pedagogical knowledge and case-level teaching exemplars, and it links them so that retrieved principles are grounded in similar student mistakes and teacher responses. Traditional RAG retrieves standalone text chunks and does not explicitly model these cross-source links, which may limit evidence integration for actionable feedback. MisEdu-RAG addresses this by organizing domain references and real cases into two-layer hypergraphs and retrieving evidence to support misconception retrieval and instructional response generation. Our study aims to answer \textbf{R}esearch \textbf{Q}uestions (\textbf{RQs}): 

\begin{itemize}
    \item[-] \textbf{RQ1:} To what extent does MisEdu-RAG improve response generation and retrieval compared to baselines, and which components contribute most?
    \item[-] \textbf{RQ2:} How well does MisEdu-RAG align with the novice teachers' needs, and to what extent do its outputs support its instructional usability?
\end{itemize}

We examine these questions on \textit{MisstepMath} \cite{ansari2025misstepmath}, a diverse student misconception dataset for math teacher training, observing significant improvements in both retrieval precision and instructional response quality over strong baselines, including LLM generation \cite{hurst2024gpt}, StandardRAG \cite{zhang2024raft}, and HyperGraphRAG \cite{feng2025hyper}. Our Contributions are as follows.
\begin{itemize}
  \renewcommand\labelitemi{$\bullet$} % •
  \item We propose a dual-hypergraph RAG method in Education to assist novice math teachers in solving students' misconceptions.
    \item Our qualitative analysis indicates the need for developing AI-assisted systems that help novice teachers address student misconceptions and meet their demand for timely instructional support.
  \item We construct a benchmark that organizes a concept hypergraph of pedagogical knowledge with an instance hypergraph of student mistakes to evaluate retrieval and instructional response generation.
  \item We provide systematic evidence that MisEdu-RAG yields consistent gains across LLM backbones and show that concept layer mainly improves strategy diversity and empowerment, while instance layer provides concrete guidance.

\end{itemize}

\section{Related Work}
\subsection{Professional Development for Novice Teachers}
Professional development (PD) aims to strengthen teachers' professional competencies and to better prepare them to effectively integrate emerging AI tools into classroom practice \cite{nagae2025effects}. In real classrooms, one recurring challenge is diagnosing and addressing student misconceptions, which demands strong pedagogical knowledge. For novice teachers, PD on misconception handling helps them diagnose students' mistakes and choose actionable instructional strategies. Evidence from novice elementary math teachers reveals a clear gap between recognizing common misconceptions and providing effective responses in practice \cite{moosapoor2023new}. This gap persists because novice teachers often struggle to translate what they notice into the interpretation of students' thinking and teaching strategies \cite{arslan2025study}. Providing timely and individualized support across diverse misconception types is resource intensive and hard to sustain. This limitation has motivated researchers to design AI tools that provide on-demand support for novice teachers.

\subsection{LLM-based Support for Novice Teachers}
Recent improvements in LLM performance have enabled broader use of LLMs in educational settings and research. Recent research has shifted focus from direct student tutoring to empowering novice teachers, particularly in identifying and addressing student misconceptions \cite{moosapoor2023new}. Prior work reports that novice teachers often struggle to diagnose the actual reasons for student misconceptions because they lack specialized training of professional educators in interacting with students \cite{wang2024bridging}. LLM-based tools are increasingly being used to support novice teachers in a range of instructional activities, including lesson and curriculum planning, generating teaching suggestions, and providing guidance to solve students' mistakes \cite{barno2024scaling,divjak2025learning,faraji2025designing,lin2025automatic,lin2025can}.

In classroom use, a key concern is the reliability of AI-generated support, meaning whether the content is accurate, pedagogically appropriate and aligned with current curricula. LLMs may produce incorrect or misleading content due to hallucinations, and their static knowledge may lag behind the latest curricular updates or scientific advancements \cite{ji2023survey,zhang2024comprehensive}. To mitigate this risk, teacher-facing tools should ground their response in verified instructional materials and make the supporting evidence explicit. This motivates us to adopt RAG, which retrieves relevant evidence from a trusted corpus and conditions generation on it to improve accuracy and reduce unsupported claims \cite{lewis2020retrieval,wang2024large}.

\subsection{Graph-based Retrieval-Augmented Generation for LLMs}

RAG augments LLMs with external evidence by retrieving relevant context from a database and generating responses conditioned on it~\cite{lewis2020retrieval,li2025retrieval}. However, passage-level retrieval can become inefficient when the required evidence is scattered across multiple sources, because it often retrieves many loosely related passages, leading to redundant context and incomplete evidence coverage \cite{abdelmagied2025leveraging,han2025hypergraph}. Graph-based RAG addresses this problem by organizing corpus chunks into a structured index, where nodes represent retrievable units and edges encode their relations, enabling the retriever to return a connected evidence subgraph for coherent aggregation. GraphRAG constructs a corpus graph where nodes represent entities (namely domain concepts or keywords), and edges capture relations mined from the corpus, and it uses graph communities to build query context and improve evidence aggregation \cite{edge2024local}. Hypergraph-based RAG further extends graph-based RAG by using hyperedges to connect multiple nodes, which captures multiple relations that are common when evidence must be combined across several passages \cite{feng2025hyper,hu2026cog}. Building on these approaches, MisEdu-RAG supports misconception-centered novice math teacher training by building a concept hypergraph from pedagogical references and an instance hypergraph from student mistake cases, then using two-stage retrieval to link principles to similar cases. It helps teachers turn students’ misconceptions into actionable teaching strategies.

\section{MisEdu-RAG}
\begin{figure}[h]
\centering
\includegraphics[width=0.92\textwidth]{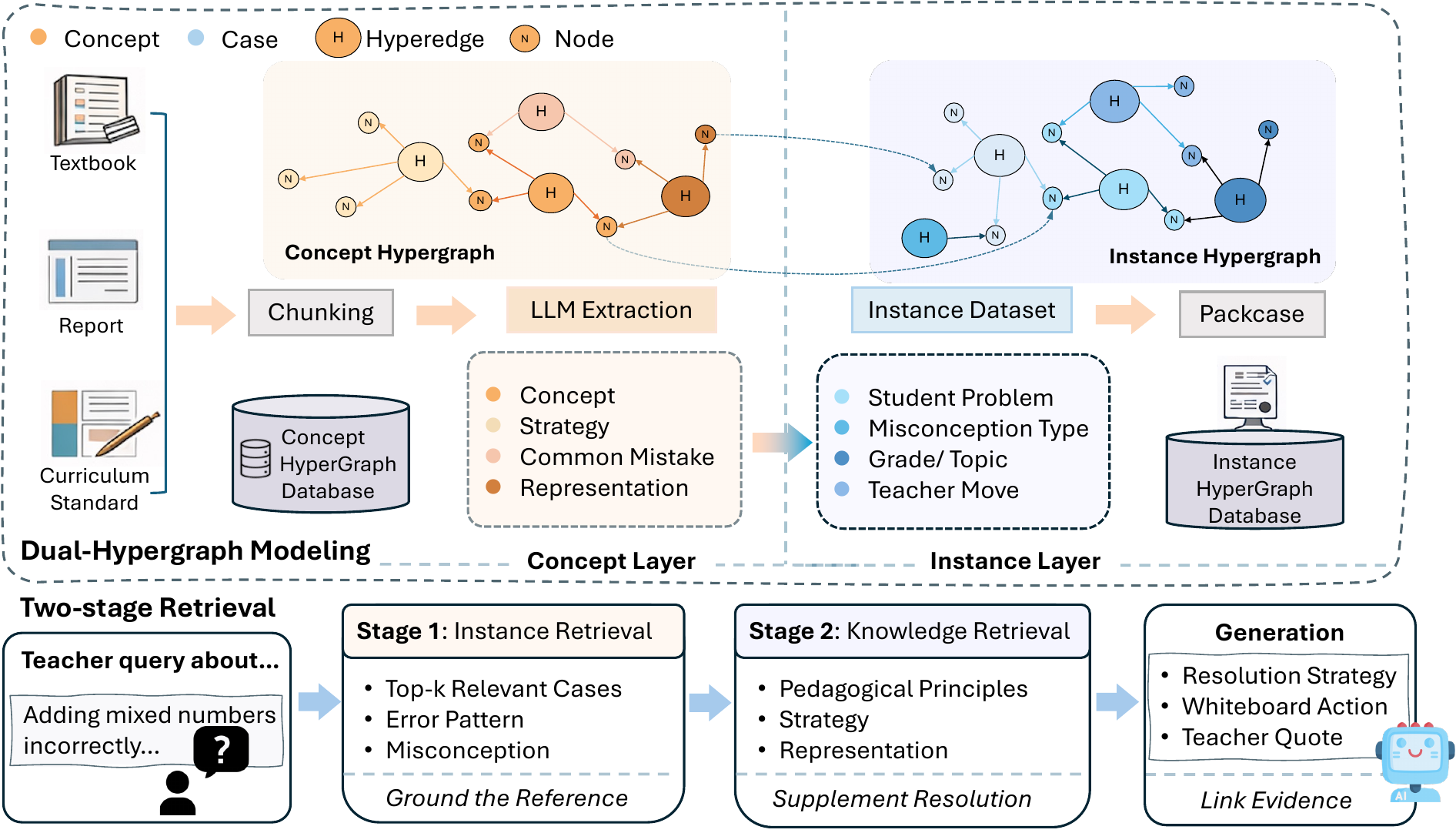}
\caption{Schematic diagram of our proposed MisEdu-RAG framework.}
\label{fig3}
%Challenge types are ranked by teachers' reported need.
\vspace{-5mm}
\end{figure}

\vspace{-1mm}
In this section, we provide a comprehensive overview of our proposed MisEdu-RAG model, including the process of knowledge extraction and two-stage retrieval. As shown in Figure \ref{fig3}, MisEdu-RAG organizes pedagogical knowledge and student cases into a two-layer hypergraph to support misconception retrieval and instructional response generation.
\vspace{-3mm}

%------------Figure

\vspace{-3mm}
\subsection{Preliminaries}
The hypergraph $\mathcal{H}=\{\mathcal{V},\mathcal{E}\}$ \cite{feng2019hypergraph,xue2026role} is a generalized graph, where $\mathcal{V}$ denotes the entity set and $\mathcal{E}$ is the hyperedge set. Each hyperedge can connect two or more entities, including pairwise low-order associations and high-order correlations.

\vspace{-3mm}
\subsection{Dual-Hypergraph Modeling}
To support misconception retrieval and response generation, MisEdu-RAG organizes multi-source, multi-type educational evidence into a dual-hypergraph. The two hypergraphs are complementary, with a knowledge-driven concept hypergraph constructed from pedagogical references (e.g., curriculum concepts and instructional principles grounded in learning science) and a data-driven instance hypergraph constructed from student mistake cases. 
This two-layer structure enables the two-stage pipeline that first retrieves similar cases to ground the query and then retrieves concept-level evidence to guide generation.

\vspace{-3mm}
\subsubsection{Knowledge-Driven Concept Hypergraph.}
The goal of knowledge extraction is to transform raw educational documents (e.g., \textit{Adding It Up: Helping Children Learn Mathematics} \cite{findell2001adding} and \textit{Principles to Actions: Ensuring Mathematical Success for All} \cite{leinwand2014principles}) into a structured concept database, which supports efficient and precise retrieval
\cite{li2025retrieval}. In MisEdu-RAG, we construct a knowledge base containing mathematics textbooks, curriculum standards, and research reports. It models universal educational concepts and their relationships, which remain valid across various teaching scenarios.

We extract instructional concepts as entities and represent sentences that connect multiple concepts as $n$-ary hyperedges. Specially, given a concept corpus $\mathcal{D}_{\text{con}}$, we obtain an entity set $\mathcal{V}_{\text{con}}$ and a set of concept hyperedges $\mathcal{E}_{\text{con}}$ via prompted extraction:
\begin{equation}
\left\{
\begin{aligned}
& \mathcal{V}_{\text{con}} = \text{LLM}\!\left(\mathcal{P}_{\text{con\_enti}}(\mathcal{D}_{\text{con}})\right), \\
& \mathcal{E}_{\text{low}} = \text{LLM}\!\left(\mathcal{P}_{\text{con\_low}}(\mathcal{D}_{\text{con}}, \mathcal{V}_{\text{con}})\right), \\
& \mathcal{E}_{\text{high}} = \text{LLM}\!\left(\mathcal{P}_{\text{con\_high}}(\mathcal{D}_{\text{con}}, \mathcal{V}_{\text{con}})\right).
\end{aligned}
\right.
\label{eq:concept_extraction}
\end{equation}
where $\mathcal{P}_{\text{con\_enti}}$, $\mathcal{P}_{\text{con\_low}}$, and $\mathcal{P}_{\text{con\_high}}$ are prompts for entity extraction, low-order relation extraction, and high-order relation extraction, respectively.
$\mathcal{E}_{\text{low}}$ captures low-order (pairwise) relations between concept entities, while $\mathcal{E}_{\text{high}}$ represents high-order relations that involve multiple entities within the same instructional statement. The resulting concept hypergraph $\mathcal{H}_{\text{con}}=\{\mathcal{V}_{\text{con}}, \mathcal{E}_{\text{low}}, \mathcal{E}_{\text{high}}\}$ serves as the pedagogical knowledge scaffold.

\vspace{-3mm}
\subsubsection{Data-Driven Instance Hypergraph.}
While the concept layer provides general pedagogical principles, novice teachers also need concrete guidance that reflects real-world teaching scenarios. Therefore, we construct an instance hypergraph $\mathcal{H}_{\text{ins}}$ from student misconception instances $\mathcal{D}_{\text{ins}}$ derived from \textit{MisstepMath} records. Each case contains structured fields such as grade, topic, challenge type, the student's mistake description, and the teacher's resolution.
\begin{equation}
\left\{
\begin{aligned}
& \forall d_i \in \mathcal{D}_{\text{ins}},\ \mathcal{K}_i = \text{Keys}(d_i),\\
& \mathcal{V}_{\text{ins}}^{i} = \{(k, v_{i,k}) \mid k \in \mathcal{K}_i,\ v_{i,k}=\text{Val}(d_i,k)\}, \\
& \mathcal{E}_{\text{ins}}^{i}=\text{PackCase}(\mathcal{V}_i).
% & \mathcal{H}_{\text{ins}}=(\mathcal{V}_{\text{ins}},\mathcal{E}_{\text{ins}}).
\end{aligned}
\right.
\label{eq:instance_hypergraph}
\end{equation}
As shown in Eq.~\ref{eq:instance_hypergraph}, for each case $d_i$, we first extract its field keys $\mathcal{K}_i=\text{Keys}(d_i)$. We then construct the instance entity set $\mathcal{V}_{\text{ins}}^{i}$ by pairing each key $k \in \mathcal{K}_i$ with its corresponding value $v_{i,k}=\text{Val}(d_i,k)$. Finally, $\text{PackCase}(\cdot)$ serializes each case entity set $\mathcal{V}_{\text{ins}}^{i}$ into a single retrievable case-level hyperedge $\mathcal{E}_{\text{ins}}^{i}$. 
After processing all cases, we aggregate all entities into 
$\mathcal{V}_{\text{ins}}$ and all case hyperedges into $\mathcal{E}_{\text{ins}}$. 
The resulting instance hypergraph is $\mathcal{H}_{\text{ins}}=(\mathcal{V}_{\text{ins}},\mathcal{E}_{\text{ins}})$, where $\mathcal{V}_{\text{ins}}$ denotes the instance entity set and $\mathcal{E}_{\text{ins}}$ represents the set of case-level hyperedges.

\subsection{Two-Stage Retrieval}
Motivated by the cognitive process of expert teachers, who typically resolve problems by first recalling similar past experiences and then verifying them against pedagogical theories, we design a cognitive-inspired two-stage retrieval strategy. The first stage retrieves similar cases as references, while the second stage retrieves concept‑level knowledge to refine and supplement the initial response. The two stages are complementary and jointly support more instructive outputs.

\vspace{-3mm}
\subsubsection{Instance Retrieval Reference.}
This stage aims to retrieve concrete student cases from the instance hypergraph $\mathcal{H}_{\text{ins}}$ as contextual references, mimicking the cognitive process of recalling similar teaching experiences. 

Specifically, for a given user query $\mathcal{Q}$, we extract query keywords to perform semantic matching via a retrieval function $\mathcal{R}$ against the instance hypergraph database. This process retrieves similar instance content to serve as references, ultimately selecting the top-$k$ relevant instance hyperedges, denoted as $\mathcal{E}_{\text{ret}}$. 
To capture additional reference information with structured context, we perform a neighbor expansion process on the instance hypergraph to retrieve the associated vertex entities $\mathcal{V}_{\text{ctx}}$ connected to these hyperedges. Finally, the query and the retrieved case context are fed into the LLM to generate a preliminary, instance-grounded response $\mathcal{A}_{\text{ins}}$:
\begin{equation}
\left\{
\begin{aligned}
& \mathcal{E}_{\text{ret}} = \mathcal{R}(\mathcal{P}_{\text{key}}(\mathcal{Q}), \mathcal{E}_{\text{ins}}), \\
& \mathcal{V}_{\text{ctx}} =  \mathcal{N}(e_{\text{ret}}, \mathcal{H}_{\text{ins}}), e_{\text{ret}}\in \mathcal{E}_{\text{ret}}, \\
& \mathcal{A}_{\text{ins}} = \text{LLM}(\mathcal{Q}, \mathcal{E}_{\text{ret}}, \mathcal{V}_{\text{ctx}}, \mathcal{D}_{\text{ins}}).
\end{aligned}
\right.
\label{eq:retrieval_diffusion}
\end{equation}

\vspace{-5.5mm}
\subsubsection{Knowledge Retrieval Supplement.}
Although instance retrieval offers empirical evidence from historical cases, it may lack theoretical depth and generalized pedagogical principles. To address this, the second stage leverages the concept hypergraph $\mathcal{H}_{\text{con}}$ to provide a theoretical supplement to the initial response $\mathcal{A}_{\text{ins}}$.

Specifically, we treat the generated instance-based answer $\mathcal{A}_{\text{ins}}$ and user query $\mathcal{Q}$ as the concept query source $\mathcal{Q}_{\text{con}}$. We extract key pedagogical concepts from $\mathcal{Q}_{\text{con}}$ and retrieve similar entities from the entity set $\mathcal{V}_{\text{con}}$ within the concept hypergraph. By traversing the relations connected to these concepts, we retrieve a context concept subgraph $\mathcal{G}_{\text{ctx}}$ that explains the underlying logic of the student's mistake. The final instructional response is generated by synthesizing the user query, the instance-based response, and the retrieved theoretical knowledge:
\begin{equation}
\left\{
\begin{aligned}
& \mathcal{V}_{\text{ret}} = \mathcal{R}(\mathcal{P}_{\text{con}}(\mathcal{Q}_{\text{con}}), \mathcal{V}_{\text{con}}), \\
& \mathcal{G}_{\text{ctx}} = \mathcal{N}(v_{\text{ret}}, \mathcal{H}_{\text{con}}), v_{\text{ret}} \in \mathcal{V}_{\text{ret}}, \\
& \mathcal{A}_{\text{final}} = \text{LLM}(\mathcal{Q}, \mathcal{A}_{\text{ins}}, \mathcal{G}_{\text{ctx}}, \mathcal{D}_{\text{con}}).
\end{aligned}
\right.
\label{eq:knowledge_retrieval}
\end{equation}
where $\mathcal{P}_{\text{con}}$ denotes the concept extraction prompt.

\vspace{-3mm}
\section{Study Setup}
\vspace{-1mm}

\subsubsection{Datesets.} We evaluate MisEdu-RAG on \textit{MisstepMath}\footnote{\url{https://huggingface.co/datasets/LLMEducation/MisstepMath}}, a semi-synthetic dataset spanning from Kindergarten to Grade 8 and covering a wide range of mathematics topics and sub-topics \cite{ansari2025misstepmath}. It contains 12,000 categorized student mistake cases paired with instructional teacher responses. Each case is annotated with mistake categories such as conceptual misunderstandings, procedural mistakes, and language-related barriers. We use \textit{MisstepMath} to construct the instance layer and to derive evaluation queries for retrieval and feedback generation. For the concept layer, we compile a pedagogy-oriented corpus from authoritative mathematics education resources, including curriculum standards, research reports and evidence-based intervention guidelines \cite{core2010common,findell2001adding,fuchs2021assisting,leinwand2014principles}. 

\vspace{-3mm}
\subsubsection{Baselines.} We compare MisEdu-RAG with three baselines: LLM generation \cite{hurst2024gpt}, which directly uses an LLM to generate feedback without retrieval; StandardRAG \cite{zhang2024raft}, which retrieves top-ranked text chunks from the same corpus and then generates; and HypergraphRAG \cite{feng2025hyper}, a hypergraph-based retriever that aggregates relational evidence before generation.

\vspace{-3mm}
\subsubsection{Evaluation Metrics.} 

We evaluate retrieval precision with cosine similarity and token-level F1 scores and adopt a scoring-based assessment as the primary metric for evaluating response generation.

\vspace{-3mm}
\paragraph{Retrieval Metrics.} We report cosine similarity and token-level F1 scores to capture semantic and lexical match between retrieved text and gold answers. % Cosine similarity measures embedding-level semantic agreement, while F1 scores measure overlap at the token level. 
Let $\hat{e}$ denote the retrieved text, and let $\mathcal{E}$ denote the set of acceptable golden answers. We compute cosine similarity as

\begin{equation}
\mathrm{Cosine}(\hat{e}, \mathcal{E})=\max_{e \in \mathcal{E}}
\frac{f(\hat{e})^{T} f(e)}{\|f(\hat{e})\|\,\|f(e)\|},
\label{eq:cosine}
\end{equation}
where $f(\cdot)$ is a fixed text embedding function. The dataset-level cosine score is the arithmetic mean over all examples.

\noindent\emph{Token-level F1 scores} is computed after normalization $N(\cdot)$ and tokenization $\tau(\cdot)$. Let $\tilde{T}=\tau(N(\hat{e}))$ and $T=\tau(N(e))$ denote the resulting token multisets for the retrieved text and a golden answer, respectively. The metric is defined as
\begin{equation}
\mathrm{F1}(\hat{e}, \mathcal{E})=\max_{e \in \mathcal{E}}
\frac{2\sum_{w \in V}\min\!\left(c_{w}(\tilde{T}),\,c_{w}(T)\right)}
{|\tilde{T}|+|T|},
\label{eq:tokenf1}
\vspace{-1mm}
\end{equation}
where $V$ is the union of tokens in $\tilde{T}$ and $T$, $c_{w}(\cdot)$ counts token multiplicities, and $|\cdot|$ denotes token length. We set $\mathrm{F1}=1$ if both sides are empty and $\mathrm{F1}=0$ if only one side is empty. We report dataset-level F1 as the mean over examples.

\vspace{-3mm}
\paragraph{Generation Metrics.}
We evaluate instructional response generation quality using a scoring-based assessment that better reflects pedagogical quality in open-ended settings \cite{feng2025hyper}. Since this method depends on the reference, we follow prior work and use the source chunk from which each question is derived as the reference evidence. Then we prompt the judge LLM to score each response on five dimensions that capture teaching quality criteria for misconception feedback: covering key ideas, offering alternative strategies, empowering student reasoning, maintaining logical coherence, and being easy to understand, corresponding to \emph{Comprehensiveness}, \emph{Diversity}, \emph{Empowerment}, \emph{Logicality}, and \emph{Readability}. 

Let $\hat{y}$ denote a generated response and let $s_d(\hat{y})$ be its score on dimension $d \in \mathcal{D}$, where $\mathcal{D}=\{\mathrm{Comp}, \mathrm{Div}, \mathrm{Emp}, \mathrm{Log}, \mathrm{Read}\}$. \textit{Comp.} for Comprehensiveness, \textit{Div.} for Diversity, \textit{Emp.} for Empowerment, \textit{Log.} for Logical, and \textit{Read.} for Readability. We compute the overall score as the mean across the five dimensions:
\begin{equation}
\mathrm{Overall}(\hat{y})=\frac{1}{|\mathcal{D}|}\sum_{d \in \mathcal{D}} s_d(\hat{y}).
\label{eq:overall_score}
\vspace{-1mm}
\end{equation}

We report dataset-level scores by averaging $\mathrm{Overall}(\hat{y})$ over all examples, and also report dimension-wise averages for fine-grained analysis. Higher scores indicate greater accuracy in responses and a lower probability of hallucination.

\vspace{-3mm}
\subsubsection{Implementation Details.} We use the GPT-4o-mini model for generation with a decoding temperature of 0.2. All experiments are run on a cloud service equipped with one RTX 3090 (24GB) GPU and 90GB RAM. All compared methods share the same generator and decoding configuration.

\vspace{-3mm}
\subsubsection{Qualitative Analysis.} To evaluate the practical potential of MisEdu-RAG, we conducted a pilot study using online questionnaires and semi-structured interviews among in-service K–12 math teachers. Our current study obtained ethical approval from an [Anonymized] institution. Through snowball sampling, we collected 221 valid questionnaires, including 31.2\% novice teachers ($\leq$3 years of experience). From this pool, we selected 6 teachers for in-depth interviews through maximum variation sampling to capture needs and perceptions. Participants included teachers across elementary to high school levels and from multiple regions, including North America (N = 8) and Asia (N = 4).

\vspace{-3mm}
\section{Results}
\vspace{-1mm}

\subsection{Results on RQ1: Comparative Experiment and Ablation Study}

\subsubsection{Retrieval Task.} Table \ref{tab:baseline} compares our \textbf{MisEdu-RAG} with LLM direct generation, StandardRAG, and HypergraphRAG on various math topics, using four models: \texttt{Qwen-Plus}, \texttt{GPT-4o-mini}, \texttt{DeepSeek-R1}, and \texttt{LLaMa3.3-70B} for response generation. We observe that MisEdu-RAG achieves the best results across all four backbones, indicating that its retrieval improvements are consistent and do not depend on a specific generator model.

% --------------------Table
\begin{table}[h]
\vspace{-3mm}
\caption{Retrieval performance (Cosine, F1) of baselines under four LLM backbones. The largest value in each column is in bold.}
\label{tab:baseline}
\centering
\scriptsize
\setlength{\tabcolsep}{3.5pt}
\renewcommand{\arraystretch}{1.05}

\resizebox{\linewidth}{!}{%
\begin{tabular}{lcccccccc}
\toprule
\textbf{Method} &
\multicolumn{2}{c}{\textbf{Qwen-Plus}} &
\multicolumn{2}{c}{\textbf{GPT-4o-mini}} &
\multicolumn{2}{c}{\textbf{DeepSeek-R1}} &
\multicolumn{2}{c}{\textbf{LLaMa3.3-70B}} \\
\cmidrule(lr){2-3}\cmidrule(lr){4-5}\cmidrule(lr){6-7}\cmidrule(lr){8-9}
& \textbf{Cosine} & \textbf{F1}
& \textbf{Cosine} & \textbf{F1}
& \textbf{Cosine} & \textbf{F1}
& \textbf{Cosine} & \textbf{F1} \\
\midrule
LLM generation & 0.5612 & 0.1837 & 0.5770 & 0.1655 & 0.4766 & 0.1323 & 0.4310 & 0.1274 \\
StandardRAG     & 0.5805 & 0.1922 & 0.6136 & 0.2398 & 0.4836 & 0.1382 & 0.4780 & 0.1733 \\
HypergraphRAG   & 0.6246 & 0.2870 & 0.6202 & 0.2801 & 0.5377 & 0.1791 & 0.5026 & 0.1620 \\
\textbf{MisEdu-RAG (ours)}
               & \textbf{0.6749} & \textbf{0.3965}
               & \textbf{0.6483} & \textbf{0.3437}
               & \textbf{0.5947} & \textbf{0.2548}
               & \textbf{0.5633} & \textbf{0.2317} \\
\bottomrule
\end{tabular}}
\end{table}

Compared with HypergraphRAG, MisEdu-RAG achieves higher alignment with the reference evidence, improving embedding-space cosine similarity by $+2.81\%$ to $+6.07\%$ and token-level F1 overlap by $+6.36\%$ to $+10.95\%$ across four backbones. The gains are most notable on \texttt{Qwen-Plus} ($+5.03\%$ Cosine, $10.95\%$ F1) and \texttt{DeepSeek-R1} ($+5.70\%$ Cosine, $+7.57\%$ F1), indicating that better retrieval yields contexts that better cover the reference content, which helps the generator match the golden answers more closely, even under weaker backbones. For LLM generation and StandardRAG, MisEdu-RAG shows more significant improvements. These consistent advantages across backbones demonstrate model-agnostic performance and confirm that the two-stage retrieval design enhances evidence alignment in both embedding and lexical spaces.

\vspace{-4mm}
\subsubsection{Generation Task.} 
\begin{figure}[t]
\centering
\includegraphics[width=0.95\textwidth]{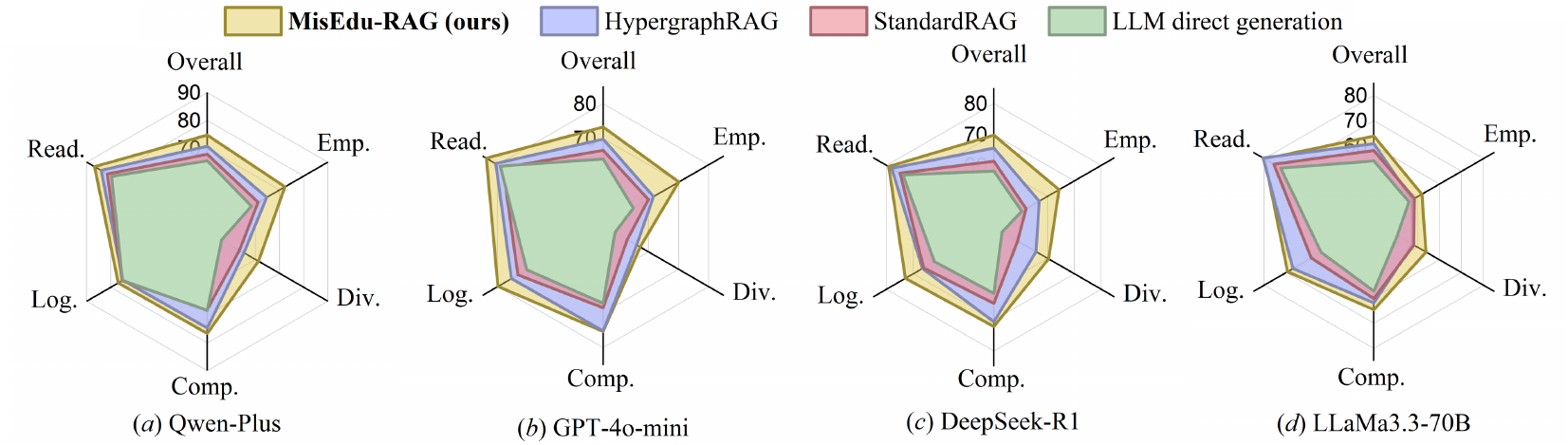}
\caption{Comparison of five-dimensional response generation quality among four generated models. 
(a) Qwen-Plus, (b) GPT-4o-mini, (c) DeepSeek-R1, (d) LLaMa3.3-70B.} 
\label{fig1}
%\vspace{-6mm}
\end{figure}

We evaluate instructional response generation using a scoring-based assessment with five distinct metrics. For a fair comparison, we randomly sample 100 questions spanning K-8 and evaluate all LLM backbones and augmentation strategies on the same query set. Fig. \ref{fig1} summarizes the results in a radar plot that jointly visualizes the overall score and detailed dimension scores. 
% Subfigures (a) to (d) correspond to Qwen-Plus, GPT-4o-mini, DeepSeek-R1, and LLaMa3.3-70B, respectively.
%\vspace{-4mm}

Fig. \ref{fig1} shows that MisEdu-RAG achieves the strongest overall performance across backbones, indicating that its retrieval design yields consistent improvement under different generators. Notably, MisEdu-RAG enhances performance by 15.32\% relative to \texttt{DeepSeek-R1} and by 12.41\% in comparison to \texttt{Qwen-Plus}, underscoring the value of constructing a hypergraph-based concept knowledge base for elevating the quality of LLM outputs. Both the baseline LLM and RAG approaches demonstrate high scores in \textit{Log.} and \textit{Read.}, which reflects that most methods can produce fluent responses due to large-scale pretraining. The largest improvements concentrate on \textit{Div.} and \textit{Emp.} because misconception retrieval provides more alternative teaching strategies and more actionable guidance, with other metrics determined by the backbone's language competence.

%\vspace{-4mm}
\subsubsection{Ablation Study.} To quantify the contribution of each hypergraph layer defined in Section~3.1, we conduct an ablation study by removing either the concept layer or the instance layer from the original MisEdu‑RAG. As shown in Table \ref{tab:ablation_cond_metrics}, we notice that MisEdu‑RAG consistently outperforms both single‑layer models on retrieval and generation tasks. In the retrieval task, MisEdu‑RAG achieves the highest Cosine and F1 score, indicating better alignment with the evidence in embedding and lexical spaces when concept and case are combined. In the generation task, MisEdu‑RAG also shows clear improvement across all five dimensions, with the largest improvements on \textit{Emp.} and \textit{Read.}, suggesting that joint evidence leads to more actionable and clearer responses. Compared with the concept‑only setting, removing case evidence reduces overall quality by 6.31 points, while removing concept evidence causes a larger drop, especially in \textit{Div.} and \textit{Emp}. These results confirm that the two layers are complementary. The concept layer contributes theoretical methods, while the instance layer provides concrete teaching guidance.

\begin{table}[t]
\caption{Ablation study of MisEdu-RAG components on misconception retrieval tasks and response generation tasks. 
% Comp. for Comprehensiveness, Div. for Diversity, Emp. for Empowerment, Log. for Logical, and Read. for Readability. 
（GPT-4o-mini）}
\label{tab:ablation_cond_metrics}
\centering
\scriptsize
\setlength{\tabcolsep}{3.5pt}
\renewcommand{\arraystretch}{1.05}

\resizebox{\linewidth}{!}{%
\begin{tabular}{lcccccccc}
\toprule
\textbf{Ablation Setting} &
\multicolumn{2}{c}{\textbf{Retrieval task}} &
\multicolumn{6}{c}{\textbf{Generation task}} \\
\cmidrule(lr){2-3}\cmidrule(lr){4-9}
& \textbf{Cosine} & \textbf{F1}
& \textbf{Overall}
& \textbf{\shortstack{Comp.}}
& \textbf{Div.}
& \textbf{Emp.}
& \textbf{Log.}
& \textbf{Read.} \\
\midrule
\textbf{MisEdu-RAG (ours)} & \textbf{0.648} & \textbf{0.344} & \textbf{73.26} & \textbf{75.40} & \textbf{57.20} & \textbf{70.10} & \textbf{79.90} & \textbf{83.70} \\
Concept-only & 0.631 & 0.318 & 66.77 & 69.09 & 45.63 & 64.69 & 72.34 & 82.15 \\
Case-only & 0.637 & 0.326 & 63.09 & 68.24 & 41.67 & 57.47 & 69.76 & 78.33 \\
\bottomrule
\end{tabular}}
%\vspace{-4mm}
\end{table}

\vspace{-3mm}
\subsection{Results on RQ2: Qualitative Analysis and Visualization}

Our survey indicates that most teachers already use AI tools for teaching support (84.54\%) and many find them helpful (78.24\%). Interviewees unanimously described growing institutional encouragement, as one teacher noted, ``\textit{The school is now emphasizing the integration of AI into teaching.}'' This acceptance of AI underscores the potential utility of tools designed to address concrete instructional challenges, such as misconception handling in novice teacher training. In our pilot study, 71.67\% of novice teachers spend at least one hour per week addressing student misconceptions. One out-of-field novice described relying on revising responses through trial and error, noting that effective approaches often require repeated attempts. This aligns with prior findings that novices often struggle to diagnose misconception causes and choose appropriate strategies, partly due to limited pedagogical knowledge and explanation repertoire \cite{hobbs2026framework}. Novices further reported that during lessons they often had to ``\textit{cater to the majority}'' and ``\textit{deliver the textbook content as is},'' leaving little capacity for immediate individualized responses, consistent with cognitive overload accounts \cite{feldon2007cognitive}.

\begin{figure}[t]
\centering
\includegraphics[width=0.9\textwidth]{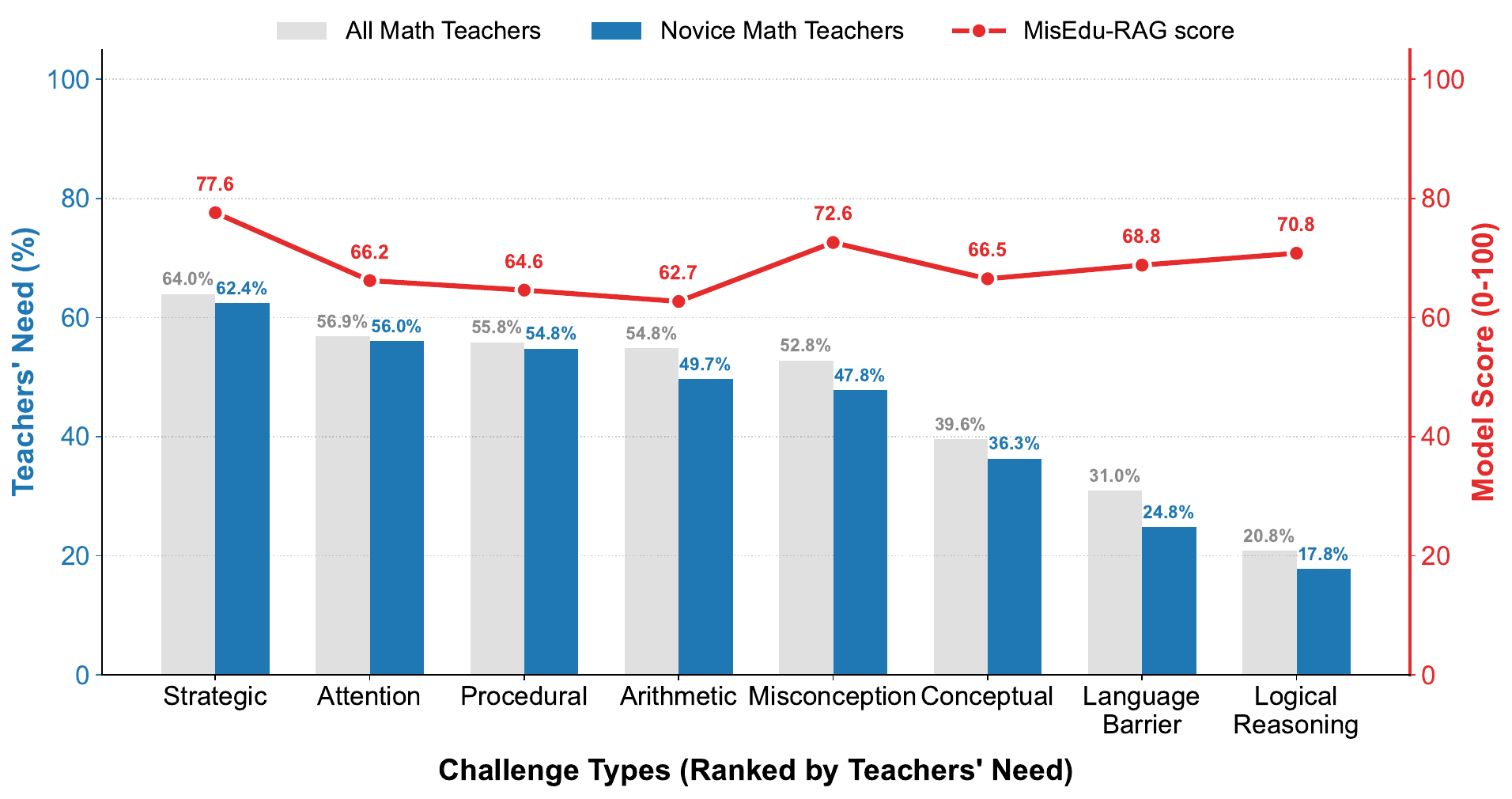}
\vspace{-2mm}
\caption{Aligning novice math teachers' instructional needs with MisEdu-RAG capability across challenge types. Bars (left) indicate prevalence (\%) for all and novice math teachers, while the line (right) shows the model's score. }
\label{fig2}
% \vspace{-3mm}
\end{figure}

The gap between students' needs and novice teachers' limited capacity highlights the need for a supportive tool, such as MisEdu-RAG. As an externalized ``\textit{cognitive partner}'' and ``\textit{experienced tutor}'', MisEdu-RAG can provide ``\textit{instant, accurate diagnostic support and step-by-step pedagogical strategy retrieval.}'' Figure~\ref{fig2} summarizes our survey results, which shows that the most frequent challenge types involve strategic problems, attention-related issues and procedural errors. MisEdu-RAG performs best on strategic challenges and remains strong on misconception and logical reasoning, aligning model capability with high-demand categories.

%------------Figure

\begin{table}[h!]
\caption{Case comparison of MisEdu-RAG and other baselines. All methods use GPT-4o-mini as the generation backbone.}
\label{tab:case_study_2x2}
\centering
\small
\setlength{\tabcolsep}{4pt}
\renewcommand{\arraystretch}{1.15}

\resizebox{\linewidth}{!}{%
\begin{tabular}{p{2.2cm} p{7.3cm} p{7.3cm}}
\toprule
\textbf{Query} &
\multicolumn{2}{p{14.6cm}}{As a teacher, how can I help students with \emph{Adding Mixed Numbers Incorrectly} when working on
$1\frac{2}{5} + 2\frac{3}{5}$? } \\
\midrule

\textbf{Golden answer} &
\multicolumn{2}{p{14.6cm}}{
TEACHER\_RESOLUTION\_STRATEGY: Teach students to handle whole numbers and fractions separately, then combine. \newline
TEACHER\_WHITEBOARD\_ACTION: Calculate $1+2$ and $\frac{2}{5}+\frac{3}{5}$ separately, then combine results. \newline
TEACHER\_QUOTE: ``Solve the whole numbers first, then the fractions.''
} \\
\midrule

\textbf{Model} & \textbf{LLM direct generation} & \textbf{StandardRAG} \\
\midrule
\textbf{Response} &
\textit{TEACHER\_RESOLUTION\_STRATEGY: Use visual aids to show the addition process step-by-step.\newline
TEACHER\_WHITEBOARD\_ACTION: Draw a number line and mark the fractions and whole numbers on it.\newline
TEACHER\_QUOTE: ``Let's break this down into simpler parts to avoid confusion.''} &
\textit{\textit{%
TEACHER\_RESOLUTION\_STRATEGY: Demonstrate how to convert mixed numbers to improper fractions before adding. \newline
TEACHER\_WHITEBOARD\_ACTION: Write the steps of converting and then adding the fractions clearly on the whiteboard. \newline
TEACHER\_QUOTE: ``Let's make sure we add the whole numbers and fractions together correctly.''%
}
} \\
\midrule
\textbf{Score} &  Cosine: 0.64, F1: 0.24, LLM\_score: 42.4  & Cosine: 0.55, F1: 0.32, LLM\_score: 53.0 \\
\midrule

\textbf{Model} & \textbf{HypergraphRAG} & \textbf{MisEdu-RAG (ours)} \\
\midrule
\textbf{Response} &
\textit{\textit{%
TEACHER\_RESOLUTION\_STRATEGY: Teach students to combine whole numbers and fractions together correctly. \newline
TEACHER\_WHITEBOARD\_ACTION: Demonstrate adding mixed numbers step-by-step on the whiteboard. \newline
TEACHER\_QUOTE: ``Remember, we have to add the whole numbers and the fractions at the same time, not separately!''%
}} &
\textit{\textit{%
TEACHER\_RESOLUTION\_STRATEGY: Guide students to combine mixed numbers turned into improper fractions before adding. \newline
TEACHER\_WHITEBOARD\_ACTION: Draw a number line showing the addition of $1\frac{2}{5}$ and $2\frac{3}{5}$. \newline
TEACHER\_QUOTE: ``Let's convert these mixed numbers to improper fractions first, then add the numerators and keep the denominator.''%
}
} \\
\midrule
\textbf{Score} & Cosine: 0.73, F1: 0.43 , LLM\_score: 67.8  & Cosine: 0.78, F1: 0.43, LLM\_score: 74.1 \\
\bottomrule
\end{tabular}%
}
\vspace{-3mm}
\end{table}

As shown in Table \ref{tab:case_study_2x2}, instructional response quality differs obviously among these methods. The table presents a query on the misconception about \textit{Adding Mixed Numbers Incorrectly} (e.g., $1\frac{2}{5}+2\frac{3}{5}$). The results show a clear performance difference: LLM direct generation produces generic guidance with limited alignment to the target misconception, StandardRAG partially improves relevance but still introduces a strategy shift that deviates from the golden answer, while HypergraphRAG further strengthens alignment and specificity. MisEdu-RAG achieves the best overall quality, suggesting that our retrieval method provides more faithful grounding and yields responses that are more consistent with expected teaching objectives. Notably, MisEdu-RAG provides clearer and more actionable steps for teaching guidance. Thus, novice teachers can easily implement it in the classroom by following the steps, allowing them to pay more attention to students and focus on teaching while gaining experience.

%------- Case study table

\vspace{-3mm}
\section{Discussion}

\vspace{-2mm}
\subsection{Implications}

Our study proposes a dual-hypergraph RAG framework to support novice math teachers in addressing student misconceptions. Our results show that MisEdu-RAG improves retrieval and instructional response generation over strong baselines, and ablations confirm that both hypergraph layers contribute meaningfully. Additionally, the qualitative analyses indicate that MisEdu-RAG can provide structured and instructional teaching guidance for students' misconception handling, with practical potential to assist novice mathematics teachers.

\vspace{-3mm}
\subsubsection{Dual-Hypergraph Two-stage Retrieval.}
Our findings provide insights for educational AI practitioners. Misconception handling benefits from structured retrieval beyond simple passage similarity, because effective support often requires connecting pedagogical principles with real student cases and combining evidence from different sources \cite{edge2024local,feng2025hyper}. MisEdu-RAG addresses this need with a dual-hypergraph and a two-stage retrieval pipeline. From a deployment perspective, the concept layer can be updated with new references, while the instance layer can expand with additional instances. This design makes the model easier to maintain and helps it remain aligned with evolving instructional needs \cite{ji2023survey}.

\vspace{-4mm}
\subsubsection{Supporting Novice Teachers in Misconception Handling.} 
Our findings contribute to practical implications for teacher professional development on misconception handling. MisEdu-RAG can support both pre-service training and early-career induction by helping novice teachers diagnose misconceptions and generate concrete teaching responses. It grounds responses in both concept-level guidance and case-level exemplars, and presents guidance in a clear format, with strategy, classroom action and teacher-facing quote. MisEdu-RAG can reduce preparation effort and ease in-class decision making for novice teachers \cite{feldon2007cognitive}. In practice, MisEdu-RAG can be used for lesson preparation, rehearsal activities and post-lesson reflection, offering timely and actionable support that fits common teacher education workflows.

\vspace{-3mm}
\subsection{Limitations}
We acknowledge several limitations of the study. First, our pilot study is based on a limited and heterogeneous sample, which may restrict generalizability. Second, \textit{MisstepMath} only focuses on U.S. K-8 mathematics misconceptions, so the grade span and pedagogical scope are constrained. Third, our current hypergraph construction is limited in capturing relations that span multiple chunks, which may lead to incomplete or fragmented knowledge signals and motivates improved cross-chunk relation fusion.

\vspace{-3mm}
\section{Conclusion}
\vspace{-2mm}
% two-stage retrieval

This work tackles the challenge novice math teachers face in addressing student misconceptions with limited teaching experience. We propose MisEdu-RAG, a dual-hypergraph RAG framework that retrieves pedagogical knowledge and relevant student cases to generate structured, actionable instructional guidance for misconception handling. Future research should explore broader subjects and age groups, and improve model construction to better capture cross-chunk relations for more complete and scalable evidence grounding.

\subsubsection{\ackname} This paper developed from a group project in EDUR8600 at The University of Hong Kong (Semester 1, 2025–2026). We thank the course instructor and our peers for their helpful feedback. This work was supported by the Faculty Research Fund and by the grant from the URC (Grant No. 2401102970) at The University of Hong Kong.

%
% ---- Bibliography ----
%
% BibTeX users should specify bibliography style 'splncs04'.
% References will then be sorted and formatted in the correct style.
%
% \bibliographystyle{splncs04}
% \bibliography{mybibliography}
%
\vspace{-3mm}
\bibliographystyle{splncs04}
\bibliography{references}

% \end{thebibliography}
% \fi
\end{document}